# PREPRINT:[1]
# Agile Development at Scale: The Next Frontier

*Torgeir Dingsøyr, SINTEF and Norwegian University of Science and Technology*
*Davide Falessi, California Polytechnic State University, San Luis Obispo, USA*
*Ken Power, Ireland*

Agile methods have transformed the way software is developed, emphasizing active end-user involvement, tolerance to change, and evolutionary delivery of products. The first special issue on agile development described the methods as focusing on "feedback and change"[1]. These methods have led to major changes in how software is developed. Scrum is now the most common framework for development in most countries, and other methods like extreme programming (XP) and elements of lean software development and Kanban are widely used. What started as a bottom-up movement amongst software practitioners and consultants has been taken up by major international consulting companies who prescribe agile development, particularly for contexts where learning and innovation are key. Agile development methods have attracted interest primarily in software engineering[1, 2] but also in a number of other disciplines including information systems[3] and project management.[4]

The agile software development methods were originally targeted towards small, co-located development teams, but are increasingly applied in other contexts. They were initially used to develop Web systems and internal IT systems, but are now used in a range of domains, including mission-critical systems. Methods that were designed for single teams of 5-9 developers have been adapted for use in projects with tens of teams, hundreds of developers, which can involve integration with hundreds of existing systems and affect hundreds of thousands of users.

Why use agile methods for large projects? Early advice from the agile community was that "scaling XP and agile projects is probably the last thing you would want to do"[5]. Advice from several fields is to reduce the size of software projects: Some see the "death of big software" as new technology allows for microservices which dramatically reduces the need for coordination[6]. Project management researchers recommend to reduce the size of projects[7] to reduce risk, and general advice in software engineering is to "simplify your product portfolio, reduce the product complexity".[8]

Although advice would be to reduce the size of software projects as much as possible, solutions often require too much work for a single team. This is often because the new solution needs to be developed fast, or that the solution is so complex or has so many dependencies to existing systems that it is deemed inefficient or impractical to split development into small projects. Large telecom products, for example, often have more than 20 teams working on the development. Agile methods are both a way to reduce risk at scale, and also to enable innovation.

So, what exactly is "large-scale agile development"? A participant in a large project expressed it as "it is like establishing a medium-size company overnight". The context of that interview was a new project involving a number of external consultants. Often, large-scale development will be in a product development setting with established teams and with established domain knowledge. There will be different needs in the different types of large-

---

scale projects. In the box we describe an example of the first type. A common definition of large-scale agile development is development efforts with more than two development teams, and such projects often have a "high number of actors and interfaces with existing systems"[9] which have implications for the development process.

Why is large-scale agile development important now? First, the global focus on digitalization has led to an increased understanding of the importance of software, how it permeates every sector of society and how it enables competitiveness and innovation. Second, early studies of agile development in large scale indicate challenges with crucial aspects, such as coordination of teams and work.[10] As new frameworks increase in popularity, there is a need for more studies. Today "few independent empirical studies exist on how the frameworks work in practice, what circumstances each framework suits best, and what the challenges are and how to overcome them".[11] Third, that company management has become more aware of the importance of software leads to a renewed focus on development method in order to ensure competitiveness. The stakes are higher now because these methods are used at larger scale.

Who is this special issue relevant for? This special issue is relevant for decision-makers at all levels, whether deciding on a framework to adopt at company or project level, or deciding on tailoring a development model or on selecting key practices for tailoring. The insight in this special issue is relevant for managers in software companies, program managers, project managers, facilitators and developers as well as people in technical roles, product owners and customer representatives. This special issue draws on studies and experience, which will complement and sometimes contradict advice from environments who have developed frameworks or own models. Methods can serve a number of functions, from supporting process improvement initiatives with the goal of improving how products and services are delivered, to making a company look attractive for partners or employees. This special issue focuses on the first function.

**Agile development and scale; The first iterations**
In the context of software development, "agile" was first used in an article in IEEE Software in 1998[12] to describe an in-house method called Agile Software Process (ASP) and Agile Software Engineering Environment (ASEE) developed at Fujitsu. Methods such as Extreme programming, Scrum, Crystal, Evo and feature-driven development followed the Dynamic software development method (DSDM) as iterative development methods.[13] Some credit Microsoft as originators of many of the work practices such as continuous build[11], while some argue that the practices have been common amongst developers since the 1960's and agile methods are "old wine in new bottles".[12] Early advice on method tailoring was to use more "disciplined" methods when many people were involved.[14]

The first wave of agile methods focused on development in teams, with emphasis on iterative development of high-priority features, prescribing few roles and easy to use artefacts to aid the development. Many companies started with long iterations and moved to shorter iterations or continuous development, many reducing "ceremony" with methods such as Kanban. When there was a need for more development teams, these would be coordinated in a separate forum where delegates from participating teams would identify and manage dependencies in tasks between the teams.

Larger development projects would often follow agile methods at team level, combined with a project management framework such as the Project Management Body of Knowledge or Prince2. An example of such a project was the 28 month and 15 million USD project for a cruise company to develop a new web-based customer booking engine.[15] Due to a high rate of changes in requirements, they decided to use agile methods in a project which was found successful by sponsors in spite of cost and schedule overruns. During the project, about 60% of requirements changed, and as an example a requirement to allow cruise passengers to choose a specific cabin instead of a class of cabins lead to a dramatic change in hospitality practice as previous systems would let users book a class and then assign cabins at check-in time. A study of the project describes the combination of structured planning using the project management framework and the iterative nature of agile methods as enabling the project to learn and adapt.[15]

In a recent historical overview of agile software development, the topic "large-scale agile" is shown to appear in the mid 2000s.[16] Ecksteins book on "*Agile Software Development in the Large*" was the first book on the topic published in 2004. The growing interest in applying agile methods in large projects is illustrated by the ranking of "burning research questions" by practitioners at the XP2010 conference which put "agile in the large" on top.[17]

A second wave of agile methods sought to address challenges of scale, replacing advice from project management frameworks on addressing layered organisations with portfolios, addressing risks, increasing the number of roles and practices for coordination and alignment across teams and in general picking up on improvement trends such as lean and adapting it to large-scale software development. Some of the new frameworks give extensive recommendations on a number of areas such as the Scaled Agile Framework, while other provide less ceremony and recommend more decision authority to autonomous teams, such as in the "Spotify"-model. So far, we have few independent studies of how these new frameworks function, and a trend to make development method a top-down decision instead of a bottom-up movement is likely to lead to challenges with adopting new practices, while on the other hand creating awareness amongst managers on the importance of development process and that process improvement comes at a cost.

**In this issue**
This special issue includes contributions on how to address main challenges which emerge when using agile methods in large software development projects and programmes:

A central question is if one should apply agile methods to large-scale development. In "Relations between Project Size, Agile Practices and Successful Software Development", Jørgensen describes a study of 196 Norwegian IT projects, which finds that projects using agile methods outperform non-agile projects also when the projects are large.

A second question is then how to adapt agile methods for large-scale development. We now have a number of new frameworks available (see box). In "Implementing Large-Scale Agile Frameworks: Challenges and Recommendations", Conboy and Carrol provide recommendations based on a study of 13 agile transformation cases over 15 years, from companies adopting frameworks such as the Scaled Agile Framework, the Spotify Model and Nexus.

Large-scale agile development will involve numerous people in many development teams. Prior studies indicate challenges with lack of alignment between teams.[18] Agile methods primarily rely on oral communication for knowledge sharing through practices such as retrospectives and pair programming. In " Spotify Guilds – How to Succeed with Knowledge Sharing in Large-scale Agile Organizations?", Smite and her co-authors provide research-based advice on success criteria for what is more commonly known as Communities of Practice.

Another challenge in large-scale development is to ensure customer collaboration as there are often numerous stakeholders. In "Tailoring Product Ownership in Large-Scale Agile", Bass and Haxby describe product owner behaviours valued by experienced product owners and line managers, based on studies of 21 organisations.

A final challenge with scale is to be efficient in decision making in large projects or product development efforts. In "Empower Your Agile Organization – Community-based Decision Making in Large-scale Agile at Ericsson", Paasivaara and Lassenius describe a mode of decision making which seeks to preserve team autonomy in a globally distributed organization with up to 40 teams working on a product.

**Future directions**
Large-scale agile development today receives widespread interest. What will we see with respect to development of method advice for large-scale agile development in the coming years? Although there are a number of frameworks presented, we believe advice on large-scale agile development are still in a nascent state. An over-reliance on frameworks can be a dangerous thing. We see evidence of people embracing frameworks for large-scale agile, without consideration of what problem they are trying to solve, and if the framework is really an aid. Like with agile itself, frameworks should never be the goal; frameworks should help achieve a goal. Introducing a scaling framework is a significant change effort in many organizations, and should be approached with care. In particular, organizations should take care to understand why something works at a smaller scale before attempting it at a larger scale. There is also the danger of people treating these frameworks as context-free recipes, and blindly following the framework without due consideration of their context. In the coming years, we need practitioners to share their experience and express needs for research areas, we need consultants to pick up research findings and to continue integrating their experience from a number of clients. Finally, we need researchers to provide contextually relevant advice by conducting empirical studies and combining lessons with prior research from relevant fields such as project management, organisational psychology and management science. It is also critical for both researchers and practitioners to understand the basic theory behind these practices, so they can better scale them. After all it is a learning process, scale raises a set of new challenges, but it is still "about feedback and change".

**Acknowledgement**
The work with this special issue was supported by the Research Council of Norway through grant 236759. Thanks to Knut Rolland at SINTEF and University of Oslo for comments on an earlier version of this introduction. We are very grateful to authors who have submitted articles and to reviewers who have contributed greatly to the result.

**References**

1. Williams, L. and Cockburn, A., "Agile Software Development: It's about Feedback and Change," *IEEE Computer,* vol. 36, pp. 39-43, 2003.
2. Dingsøyr, T., Nerur, S., Balijepally, V., and Moe, N. B., "A Decade of Agile Methodologies: Towards Explaining Agile Software Development," *Journal of Systems and Software,* vol. 85, pp. 1213-1221, 2012.  10.1016/j.jss.2012.02.033
3. Abrahamsson, P., Conboy, K., and Wang, X., ""Lots done, more to do": the current state of agile systems development research," *European Journal of Information Systems,* vol. 18, pp. 281-284, 2009.
4. Niederman, F., Lechler, T., and Petit, Y., "A Research Agenda for Extending Agile Practices In Software Development and Additional Task Domains," *Project Management Journal,* vol. 49, pp. 3–17, 2018.  https://doi.org/10.1177/8756972818802713
5. Reifer, D. J., Maurer, F., and Erdogmus, H., "Scaling agile methods," *IEEE software,* vol. 20, pp. 12-14, 2003.
6. Andriole, S. J., "The death of big software," *Communications of the ACM,* vol. 60, pp. 29-32, 2017.

7.  Flyvbjerg, B., Bruzelius, N., and Rothengatter, W., *Megaprojects and risk: An anatomy of ambition*: Cambridge University Press, 2003.
8.  Ebert, C., "50 Years of Software Engineering," *IEEE Software,* vol. 35, pp. 94-101, 2018.
9.  Rolland, K. H., Fitzgerald, B., Dingsøyr, T., and Stol, K.-J., "Problematizing Agile in the Large: Alternative Assumptions for Large-Scale Agile Development," in *International Conference on Information Systems*, Dublin, Ireland, 2016.
10. Paasivaara, M., Lassenius, C., and Heikkila, V. T., "Inter-team Coordination in Large-Scale Globally Distributed Scrum: Do Scrum-of-Scrums Really Work?," in *Proceedings of the ACM-IEEE International Symposium on Empirical Software Engineering and Measurement*, ed New York: IEEE, 2012, pp. 235-238.
11. Ebert, C. and Paasivaara, M., "Scaling agile," *IEEE Software,* vol. 34, pp. 98-103, 2017.
12. Dybå, T. and Dingsøyr, T., "Empirical Studies of Agile Software Development: A Systematic Review," *Information and Software Technology,* vol. 50, pp. 833-859, 2008. doi:10.1016/j.infsof.2008.01.006
13. Larman, C. and Basili, V. R., "Iterative and Incremental Development: A Brief History," *IEEE Computer,* vol. 36, pp. 47-56, 2003.
14. Boehm, B. and Turner, R., *Balancing Agility and Discipline: A Guide for the Perplexed*: Addison-Wesley, 2003.
15. Batra, D., Xia, W., VanderMeer, D., and Dutta, K., "Balancing Agile and Structured Development Approaches to Successfully Manage Large Distributed Software Projects: A Case Study from the Cruise Line Industry," *Communications of the Association for Information Systems,* vol. 27, p. 21, 2010.
16. Hoda, R., Salleh, N., and Grundy, J., "The Rise and Evolution of Agile Software Development," *IEEE Software,* 2018.
17. Freudenberg, S. and Sharp, H., "The Top 10 Burning Research Questions from Practitioners," *IEEE Software,* pp. 8-9, 2010.
18. Bick, S., Spohrer, K., Hoda, R., Scheerer, A., and Heinzl, A., "Coordination Challenges in Large-Scale Software Development: A Case Study of Planning Misalignment in Hybrid Settings," *IEEE Transactions on Software Engineering,* 2017. 10.1109/TSE.2017.2730870

# Perform: An Example Large-Scale Project

The Norwegian Public Service Pension Fund needed a new office automation system due to a public reform. When starting the development, the content of the reform was still to pass through parliament, and the project adopted an agile development method.

The project "Perform", was one of the largest IT projects in Norway with a final budget of about EUR 140 million. The four-year project involved 175 people, 100 were external consultants from five companies. About 800,000 person-hours were used to develop around 300 epics with a total of about 2,500 user stories. These epics were divided into 12 releases.

An existing office automation system was client/server-based and written in C. The new system was a service-oriented system written in Java. The database from the old system was kept, but the data model was changed. The regulations and legislations were implemented in the new system as rules using JRules. The system was integrated with a new document archive and with systems from another public department.

An example release contained coupling of workflow in the office automation system to an archive solution, a self-service solution for new legislation, simulation of services towards external public departments, and first data warehouse reports on new data warehouse architecture. Most user stories were identified prior to the first release but were supplemented and reprioritised for every release.

While staring small, the development project at its peak involved 12 Scrum teams working in parallel. There were numerous dependencies between work for the teams, and to ensure coordination, the teams had extra roles such as a technical architect, functional architect and a test responsible as well as several extra arenas in addition to Scrum of Scrum meetings. The product was demonstrated every three weeks after ending an iteration. Product owners were supported with extra resources, with a total of 30 people from the line organization working with defining user stories.

Key characteristics:

* 2 300 user stories
* EUR 140 million total cost
* 12 development teams
* 800,000 person-hours
* 12 releases
* 30 people from line organization involved

The organization of the program is described in a case study, which can be found here:

Dingsøyr, T., Moe, N. B., Fægri, T. E., & Seim, E. A. (2018). Exploring Software Development at the Very Large-Scale: A Revelatory Case Study and Research Agenda for Agile Method Adaptation. *Empirical Software Engineering, 23*(1), 490-520. doi:https://doi.org/10.1007/s10664-017-9524-2

# Practitioner opinion: Agility @ Scale: When a Small Cross Functional Team is Not Enough

*Steve Adolph, PhD – Senior Consultant cPrime*

It is a project manager's worst-case scenario – congestive collapse. Everyone is running around hair on fire busy, and nothing is getting done. A looming drop-dead date that is not just an arbitrary milestone, but reflects a real must-ship date or otherwise over a year's development effort will be lost. Many of the teams are so called "agile teams" but the agile magic is not happening. This is the situation I found myself in at a major industrial equipment manufacturer, some 150 plus engineers trying to fight what they believed was a battle lost. With the clock ticking, we stood the project down for two weeks, trained the teams in a scaling methodology, re-planned to create a coordinated backlog, a release plan, and relaunched. The creation of a coordinated program backlog, and bringing all teams together in a classic "big room planning" session aligned the teams and enabled them to focus on getting something done. A viable product was shipped on time. For this client, agile was no longer just a team methodology, but rather a business strategy.

Nearly 20 years ago Agile manifesto captured the imagination of many software developers burden by what Alistair Cockburn called "big-M" methodologies and the excesses of the late 90s software quality rage. Teams began experimenting with agile methodologies and liked the results. Self-reporting surveys repeatedly demonstrated agile teams created greater customer satisfaction, faster, with higher quality and lower costs. Agile demonstrability worked, and organizations wanted more of it.

As organizations demanded more, the challenge became how do we create agility beyond the team? While some teams could create end to end value on their own, the team was often just one step in an organization's value creation process. Also, that team often had to coordinate their work with other teams to create value. Just how did that supposedly omniscient product owner come up with those user stories and what happened to the "increment" after that product owner accepted it? How were the team's fast learning cycles influencing the enterprise's entire value creation process? How do we coordinate value delivery by multiple teams, and more importantly, how do we coordinate their learning? How do we manage customer needs when a user story only captured tiny sliver of value - slivers of values so thin, customers often regarded them as merely "nerd" details? Ignoring these questions or just punting the answers to some higher level product owner meant the "agile" methodologies only provided guidance for small teams, or for enterprises where teams could be organized as multiple independent, feature teams.

Some practitioners remembered agile is more than just a basket of methodologies, that it's a strategy, a mindset. That it is a competitive strategy for creating value by learning faster than the rate of change. That the economics of agility are a function of time and not size. Some agilists began exploring how that mindset could be applied to larger and more complex systems; realizing agile had not displaced 50 years of software engineering experience. In fact, they looked at how to exploit that knowledge to accelerate the learning process. They began exploring new patterns of planning, such as multi-level adaptive planning. They began integrating the concepts of both intentional and emergent architectures into new patterns of agile architecture. They looked at how to utilize frequent feedback from demos to guide the analysis process, to learn what was truly valuable and to quickly prune less valuable requirements. They looked for how to seek balance between individuals and interactions over processes and tools at scale.

The so called "scaling" frameworks, SAFe, DAD, Nexus, LeSS, etc. emerged from these patterns. These scaling frameworks integrated patterns for roles, practices, metrics, and supporting artifacts. They addressed the real concerns of practitioners and managers of large and complex systems. Requirements analysis mattered. Architecture mattered. Design mattered. Long range planning mattered. Specialization was sometimes necessary. But these could be performed in a framework that supported fast learning cycles and adaptation across a number of different time horizons.

These frameworks accelerated agile adoption among what many had derisively called "laggard" organizations. These organizations were slow to adopt agile because they simply did not believe agilists really understood their needs. They liked how the scaling frameworks enabled them to see beyond the product owner and offered models for aligning "the business" and "IT". They liked how the scaling frameworks did not just rely on the skills of omniscient product owners or even an omniscient "Chief" product owner. They liked how the frameworks help create much needed alignment by integrating the fast feedback cycles throughout the whole value creation process and not just within the software teams. Business and IT began to finally realize they were part of the same organization and had the same goals. The scaling frameworks helped organizations see agility as more than an IT or engineering cost reduction exercise.

The scaling frameworks can be fairly complex, introducing many new roles, artifacts, and practices. There are even scary echoes of the Big-M methodologies. But then product development is complex and lack of a willingness to understand the context of an organization simply means we will not have credibility with that organization. Simplicity in engineering is good, and Jim Highsmith once described agile methodologies as "barely sufficient process". But as Einstein once said "make everything as simple as possible, but no simpler" After all, the manifesto is all about balance.

# Frameworks for large-scale agile development

*Agile portfolio management:*
Suitable for: Organisations
Main characteristics: Introduces rolling planning, forecasts and dynamic management of portfolios, combined with normal agile practices such as standups and retrospectives. Establishes a set of core values which include "focus on value", "continuous review of portfolio", "demonstration of value from portfolio" and to encourage "collaboration and empowerment".
Created by: Barbara Roberts, Peter Coesmans, Peter Measey, Steve Messenger
More information: https://www.agilebusiness.org/agilepfm

*Disciplined Agile (DAD):*
Suitable for: One to many teams.
Main characteristics: A comprehensive framework combining ideas from agile development, lean software development and agile modelling. Introduces roles such as stakeholders and architectural owner, specialists, domain experts, technical experts, independent testers and integrators. 20 new roles appear with scale, such as chief architecture owner, chief product owner, communities of practice lead, and portfolio manager.
Created by: Scott Ambler, Mark Lines
More information: https://www.disciplinedagiledelivery.com

*Kanban Method:*
Suitable for: Organization of any size, from small to very large.
Main characteristics: The Kanban Method combines elements of the work of W Edwards Deming, Eli Goldratt, Peter Drucker, and Taiichi Ohno, as well as concepts such as pull systems, queuing theory, and flow. A significant differentiator from other methods is Kanban starts with where an organization is, and does not require the creation of new roles, ceremonies, or structures before getting started.
Created by: David J. Anderson
More information: https://leankanban.com/project/what-is-km/

*LeSS:*
Suitable for: 2-7 ("LeSS") or 8+ ("LeSS Huge") development teams.
Main characteristics: A "minimal extension" to Scrum to handle large-scale product development. Product owners supported by product managers and area product owners in "Huge", communities of practice for knowledge sharing and improvement across teams. Ideally most work done in the feature teams, however, sometimes a separate "undone department" to handle architecture, business analysis, QA or test.
Created by: Bas Vodde and Craig Larman
More information: https://less.works/

*Nexus:*
Suitable for: 3-9 development teams.

Main characteristics: Roles, events, artefacts, and rules that coordinate the work of approximately three to nine Scrum Teams working on a single Product Backlog to build an Integrated Increment that meets a goal. Extra meetings and roles added on to Scrum to coordinate teams. Common demonstration for all teams, same iteration length.
Created by: Ken Schwaber
More information: https://www.scrum.org/resources/nexus-guide

*Scaled Agile Framework (SAFe):*
Suitable for: From groups of 50-150, to whole organizations
Main characteristics: A comprehensive framework incorporating ideas from agile development and lean production. As of version 4.6, there are now multiple versions of SAFe, targeted at organizations and development efforts of different sizes. The versions are branded as "Essential SAFe", "Large Solution SAFe", "Portfolio SAFe", amd "Full SAFe". A central concept is the "agile release train" with five to 12 development teams which develop product increments every 8 to 12 weeks.
Created by: Dean Leffingwell
More information: https://www.scaledagileframework.com/about/

*Scrum at Scale*
Suitable for: Whole organizations
Main characteristics: Focuses on "networks" of Scrum teams. Separates responsibility for coordinating the "how" and "what" of work in organizations. A Scrum Master cycle with advice on how to coordinate Scrum teams, and a Product Owner cycle with advice on coordination of what is to be made (backlog prioritisation).
Created by: Jeff Sutherland
More information: https://www.scrumatscale.com/scrum-at-scale-guide/

*Spotify Model:*
Suitable for: Product development with many teams
Main characteristics: A snapshot of a rapidly-evolving model. Introduced the language of squads (basic unit of development), tribes (collection of squads working in related areas), chapters (people with similar skills), and guilds (organic community of interest). Different types of communities of interest or practice established across teams to ensure learning and alignment. Separate architectural roles such as system owners and a chief architect.
Created by: Henrik Kniberg and Anders Ivarsson
More information: Online article: https://blog.crisp.se/wp-content/uploads/2012/11/SpotifyScaling.pdf and videos: https://labs.spotify.com/2014/03/27/spotify-engineering-culture-part-1/

## *About the Authors*

**TORGEIR DINGSØYR** *is chief scientist at SINTEF Digital in Trondheim, Norway and an adjunct professor at the Norwegian University of Science and Technology. His main research interests are software process improvement, teamwork in software development, knowledge management and large-scale agile development. Contact him at torgeird@sintef.no*

**DAVIDE FALESSI** *is an associate professor of software engineering at California Polytechnic State University, San Luis Obispo. His main research interest is in devising and empirically assessing scalable solutions for the development of software-intensive systems. Falessi received a PhD in com- puter engineering from the University of Rome Tor Vergata. Contact him at d.falessi@gmail.com.*9

***KEN POWER** has held multiple positions in large technology organizations. His current responsibilities include leading global, large-scale engineering organization transformations. He has been working with agile and lean methods since 1999. He holds patents in virtualization and network management. His main interests include complex adaptive systems, sensemaking, flow-based development, software architecture, distributed systems, artificial intelligence, strategy, engineering management, and leadership. Contact him at [ken.power@gmail.com](mailto:ken.power@gmail.com)"*